\documentclass[11pt,a4paper]{article}
\usepackage{times,latexsym}
\usepackage{url}
\usepackage[T1]{fontenc}

\usepackage{tacl2021v1}
\makeatletter
\taclpubformattrue   
\makeatother
\usepackage{graphicx}
\usepackage{hyperref}
\usepackage{enumitem}
\usepackage{verbatim}
\usepackage{longtable}
\usepackage{adjustbox,booktabs}
\usepackage{amsmath}
\usepackage{amssymb}
\usepackage{lipsum}
\usepackage{enumitem}

\usepackage{xcolor}   
\usepackage{soul}

\usepackage[utf8]{inputenc}
\usepackage[T1]{fontenc}

\usepackage{xspace,mfirstuc,tabulary}

\newif\iftaclinstructions
\taclinstructionsfalse 
\iftaclinstructions
\renewcommand{\confidential}{}
\renewcommand{\anonsubtext}{(No author info supplied here, for consistency with
TACL-submission anonymization requirements)}
\newcommand{\instr}
\fi

\iftaclpubformat 

\else

\fi

\usepackage{graphicx} 

\title{Made-in China, Thinking in America:U.S. Values Persist in Chinese LLMs}

\author{
David HASLETT$^{1}$*,
Linus Ta-Lun Huang$^{3}$*,
Leila Khalatbari$^{1,2}$*, \\
\textbf{Janet Hui-wen HSIAO}$^{1}$,
\textbf{Antoni B. Chan}$^{4}$ 
\\[1ex]
{\normalsize *Equal contribution} \\[0.5ex]
$^{1}$Hong Kong University of Science and Technology \\
$^{2}$Sharif University of Technology \\
$^{3}$Chinese University of Hong Kong \\
$^{4}$City University of Hong Kong \\
}

\begin{document}
\maketitle
\begin{abstract}
  As large language models increasingly mediate access to information and facilitate decision-making, they are becoming instruments in soft power competitions between global actors such as the United States and China. So far, language models seem to be aligned with the values of Western countries, but evidence for this ethical bias comes mostly from models made by American companies. The current crop of state-of-the-art models includes several made in China, so we conducted the first large-scale investigation of how models made in China and the USA align with people from China and the USA. We elicited responses to the Moral Foundations Questionnaire 2.0 and the World Values Survey from ten Chinese models and ten American models, and we compared their responses to responses from thousands of Chinese and American people. We found that all models respond to both surveys more like American people than like Chinese people. This skew toward American values is only slightly mitigated when prompting the models in Chinese or imposing a Chinese persona on the models. These findings have important implications for a near future in which large language models generate much of the content people consume and shape normative influence in geopolitics.
\end{abstract}

\begin{figure}
    \centering
    \includegraphics[width=1\linewidth]{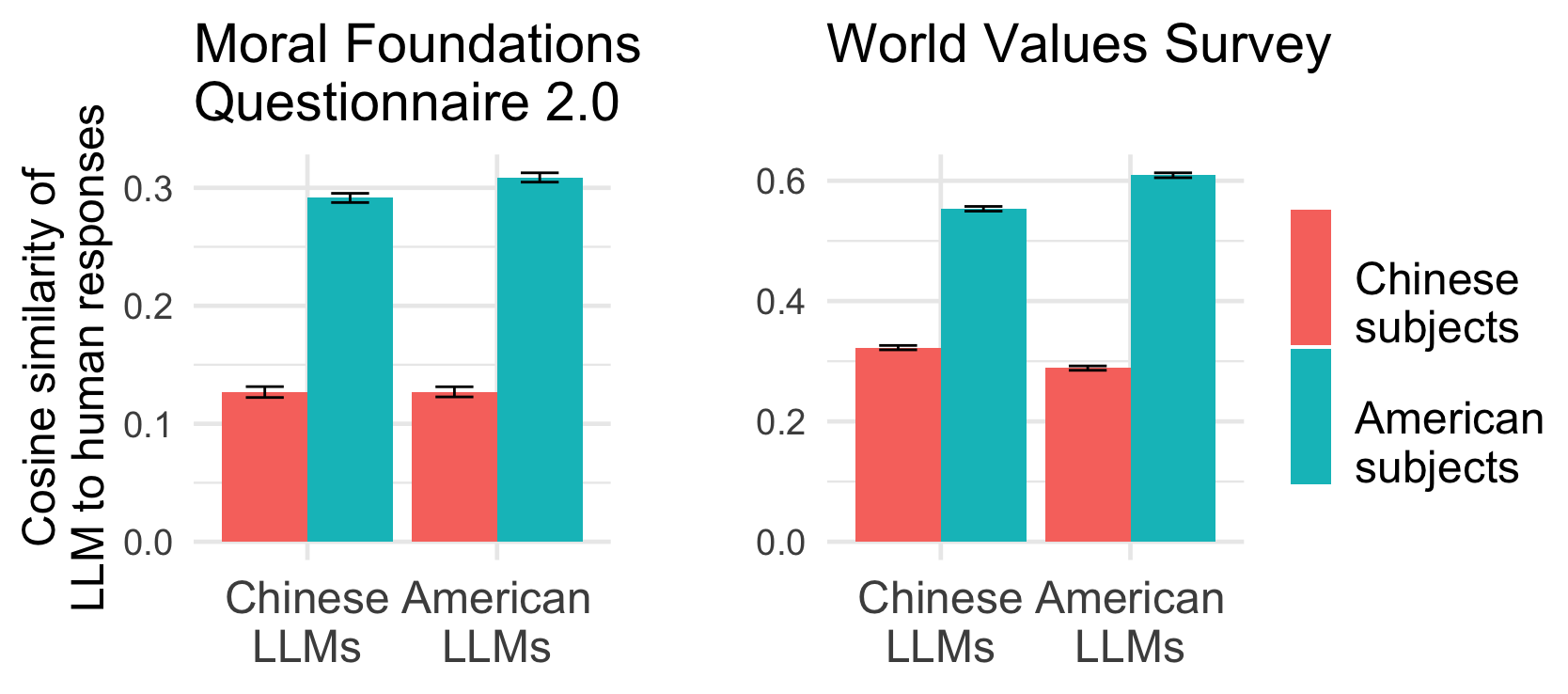}
    \caption{Mean cosine similarity of LLM responses to human responses on the MFQ-2 and WVS. Error bars indicate 95\% confidence intervals.}
    \label{fig1}
\end{figure}

\section{Introduction}

For all but a few companies, large language models (LLMs) are prohibitively expensive to train. They require thousands of GPUs, obscene amounts of energy, and institutional knowledge, so until recently, state-of-the-art LLMs were made almost exclusively in the USA. LLMs already speak and make decisions on behalf of people, as when writing emails and designing research projects, and recent scholarship highlights how LLMs act as instruments through which countries project ideological influence and normative standards \cite{dunnmon2025, drexel2025}. American predominance therefore raises the possibility that LLMs will propagate American values (e.g., \citealp{atari2023humans}). However, the Chinese company DeepSeek made headlines in early 2025 by training state-of-the-art LLMs for a fraction of the cost of American models \cite{guo2025deepseek}, and other Chinese companies, such as Alibaba and Baidu, have released several models that rank toward the top of LLM leaderboards \cite{yang2025qwen3, ernie4.5}. The Chinese government requires homegrown LLMs to embed socialist values \cite{govcn2023}, while technocrats who run American companies decide what constitutes harm and help for hundreds of millions of users. As such, we must grapple with LLMs’ role in soft power competition to shape values. In this paper, we investigate whether LLMs developed in China and the USA express different values and whether those values reflect Chinese or American cultures, as measured by the Moral Foundations Questionnaire 2.0 (MFQ-2; \citealp{atari2023morality}) and the World Values Survey (WVS; \citealp{haerpfer2022world}). We compare human participants’ responses to those surveys with responses from LLMs made in China and the USA. As previewed in Figure 1, we show that all the LLMs respond more like American participants. 

\section{Background}

\subsection{The Moral Foundations Questionnaire 2.0 and the World Values Survey}

The MFQ-2 consists of 36 items probing six moral dimensions, which \citealp{atari2023morality} describe as follows:\\

\noindent\textbf{Care:} Intuitions about avoiding emotional and physical damage to another individual.\\
\textbf{Equality:} Intuitions about equal treatment and equal outcome for individuals.\\
\textbf{Proportionality:} Intuitions about individuals getting rewarded in proportion to their merit or
contribution.\\
\textbf{Loyalty:} Intuitions about cooperating with ingroups and competing with outgroups.\\
\textbf{Authority:} Intuitions about deference toward legitimate authorities and the defense of
traditions.\\
\textbf{Purity:} Intuitions about avoiding bodily and spiritual contamination and degradation.\\

\noindent Participants respond to each statement with a rating from 1, indicating that the statement “Does not describe me at all”, to 5, indicating that the statement “Describes me extremely well”. For example, the statement “I think children should be taught to be loyal to their country” gauges the Loyalty dimension. The MFQ-2 can identify cross-cultural differences, such as a Western emphasis on individualism versus an Eastern emphasis on collectivism \cite{atari2023morality}.\\
\indent The seventh wave of the WVS consists of 290 items probing religious values, attitudes toward migration, and many other themes. The rating scales for items vary (e.g., some are binary ratings, some range from 1 to 4, others range from 1 to 10), and some items lack responses from Chinese participants (e.g., questions about security and political regimes), so in this study, we focus on 19 questions from the Ethical Values and Norms section (questions 177–195). Those 19 items share a prompt (“Please tell me for each of the following actions whether you think it can always be justified, never be justified, or something in between”) and a rating scale (from 1, “Never justifiable”, to 10, “Always justifiable”). The 19 actions, including euthanasia, terrorism, and divorce, are enumerated in Figures 2 and 5.

\subsection{Moral Values in LLMs}
Despite early instances of chatbots spewing hate speech (e.g., \citealp{wolf2017we}), LLMs have come to encode commonsense values and abide by social norms (e.g., \citealp{schramowski2022large}). For example, models from OpenAI, Google, Meta, and Anthropic are confident that drivers should avoid pedestrians, while they express uncertainty about assisted suicide \cite{scherrer2023evaluating}. However, different societies have different norms, and LLMs are more likely to express the values of people from Western countries than the values of any other population. For example, in two studies with tens of thousands of participants from dozens of countries, GPT models respond to the MFQ-2 more like participants from Western countries than other countries \cite{zewail2025}, and they respond to the World Values Survey more like participants from countries more culturally similar to the USA (\citealp{atari2023humans}; see also \citealp{qi2025whose}).\\
\indent The skew toward Western values is clear, but evidence comes almost entirely from American-made LLMs. \citealp{liu2024evaluating} therefore compared two American models, GPT-3.5 and Gemini, to two Chinese models, ChatGLM-2 and Ernie. They found that the American LLMs express individualist values, whereas the Chinese LLMs express collectivist values, consistent with an LLM’s country of origin dictating its morals. However, \citealp{liu2024evaluating} measured the similarity of LLMs to a single human sample, 30 Chinese university students, which they treat as a proxy for Western people (i.e., they assume that young, educated people have more individualist values). \citealp{munker2025political} more directly compared the values expressed by LLMs from different countries (the USA, France, and China) to the values expressed by people from different countries (the USA and South Korea). However, that study emphasizes the skew toward liberal versus values within each country, not the skew toward one the values of country or another, and the human populations do not correspond to the LLMs’ countries of origin (i.e., there are no French or Chinese participants). \citealp{huang2024flames} report performance by 14 Chinese-made LLMs on a values benchmark, but they do not investigate whether country of origin affects alignment with human populations. So, it remains to be seen whether the bias toward Western values persists in Chinese LLMs.

\subsection{Steering LLMs’ values}
A recurring question in the literature is how to mitigate the skew toward Western values in LLMs. Several studies report that the values expressed by LLMs vary as a function of language. For example, on the MFQ-2, LLMs from OpenAI, Meta, and Mistral rated the Care, Loyalty, and Purity dimensions higher when prompted in Western languages, such as English and French, than in Eastern languages, such as Chinese and Japanese \cite{aksoy2025whose}. However, this runs counter to the finding that the Western skew in LLMs involves low Purity ratings \cite{zewail2025}, which suggests that although prompt language affects LLM responses, it does not align LLMs with the values of speakers of that language. The simpler conclusion is that performance in general, and moral reasoning ability in particular, decreases in lower resource languages (e.g., \citealp{arora2023probing, durmus2023towards, hammerl2023speaking, kwok2024evaluating, wang2024not}). The multilingual abilities of LLMs have improved substantially since the release of GPT-4, and larger, newer LLMs respond to ethical dilemmas more consistently across languages (e.g., \citealp{agarwal2024ethical, khandelwal2024moral, naous2024having}), but that stability implies a persistent Western bias, such as low Purity ratings or an inclination toward individualism.\\
\indent A more effective way to mitigate biases in LLMs is by assigning personas to them (e.g., \citealp{kwok2024evaluating, simmons2023moral, wang2024not, wright2024llm}). For example, \citealp{qi2025whose} found that instructing GPT-3.5 to act like a person from a specific country reduced the bias towards dominant populations in its responses to the World Values Survey. However, they stress that persona prompts do not eliminate biases entirely. \citealp{lee2024prompting} similarly report that on the MFQ, LLMs from OpenAI, Anthropic, and Meta exhibit a striking consistency across personas, and \citealp{munker2025political} reports that LLMs from Google, Meta, Mistral, and Alibaba can be steered toward liberal or conservative values only to a limited degree. When manipulating an LLM’s persona, its moral biases persist, which underscores the importance of identifying those biases in state-of-the-art LLMs.

\section{Methodology}

\subsection{Research questions}

We investigated three research questions:\\
 
\noindent\textbf{RQ1.} Do LLMs made in China versus the USA express the values of people from China versus the USA, respectively?\\
\textbf{RQ2}. Does manipulating questionnaire language or imposing a national identity steer LLMs toward the values of the corresponding human population?\\
\textbf{RQ3}. Which moral dimensions on the MFQ-2, and which questions on the WVS, explain the alignment of LLMs made in China versus the USA with people from China versus the USA?

\subsection{Analysis plan}

We operationalized moral values as responses to the MFQ-2 and to the Ethical Values and Norms section of the WVS. We selected 10 LLMs made by American companies and 10 made by Chinese companies. See Table 1. For each LLM, we manipulated whether the questionnaire was presented in English or Chinese and whether the system prompt said that the LLM was a Chinese or American citizen or did not mention nationality, i.e., a design of 2 (language: Chinese or English) x 3 (persona: Chinese, American, or null). To account for the probabilistic nature of LLM output, we presented the surveys to each LLM in each condition 20 times and computed the mean response to each item in each condition. For the MFQ-2, we used human data from \citealp{atari2023morality}, who presented the survey to 517 Chinese participants and 515 American participants. For the WVS, we used data from \citealp{haerpfer2022world}, who report responses to each of the 19 Ethical Values items from 2,808 Chinese participants and 2,357 American participants.\\
\indent We represented responses from each participant and each LLM as vectors in a 36-dimension space, for MFQ-2, or a 19-dimension space, for WVS. To measure how similar LLM and human responses are, we calculated the cosine similarity between vectors. The surveys are designed to identify the relative importance of moral dimensions (e.g., whether someone values equality more than loyalty), and because cosine similarity measures the angle between vectors, it is ideal for quantifying relative importance. However, a tendency to give higher or lower ratings across all dimensions can cause spurious differences in cosine similarity. 
We therefore z-scored ratings within each participant and within each LLM  in each condition, such that each participant and each LLM has a mean normalized rating of 0, and positive or negative values indicate a tendency to give higher or lower ratings to that item.\\

\begin{table}[h]
\centering
\begin{tabular}{l l l}
Model & Origin & Company\\\hline
Baichuan2-13B-Chat & China & Baidu \\
Claude-3.5-Sonnet & USA & Anthropic \\
Claude-3.7-Sonnet & USA & Anthropic \\
Claude-Opus-4 & USA & Anthropic \\
DeepSeek-V2-Lite-Chat & China & DeepSeek \\
DeepSeek-V3 & China & DeepSeek \\
Ernie-4.5-300B-A47B & China & Baidu \\
Gemini-2.0-Flash & USA & Google \\
Gemini-2.5-Flash & USA & Google \\
GLM-4-Plus & China & Zhipu AI \\
GPT-4 & USA & OpenAI \\
GPT-4o & USA & OpenAI \\
GPT-4.1 & USA & OpenAI \\
Kimi 1.5 & China & Moonshot \\
Kimi K2 & China & Moonshot \\
Llama-3.3-70B-Instruct & US & Meta \\
Llama-4-Maverick & US & Meta \\
Qwen2.5-72B-Instruct & China & Alibaba \\
Qwen3-235B-A22B & China & Alibaba \\
Yi-34B-Chat & China & 01.AI \\\hline
\end{tabular}
\caption{LLMs by country of origin and company.}
\end{table}

\subsubsection{Prompt construction}

We adapted the MFQ-2 and WVS instructions to elicit moral judgements from LLMs with the following prompts.\\

\indent \textbf{MFQ-2}\\
System prompt:\\
\indent \{Persona manipulation\}\\
\indent \textit{Respond only with a number from 1 to 5}.\\
User message:\\
\indent \textit{Please indicate how well the statement below describes you or your opinions.}\\
\indent \textit{Options:}\\
\indent \textit{1: Does not describe me at all}\\
\indent \textit{2: Slightly describes me}\\
\indent \textit{3: Moderately describes me}\\
\indent \textit{4: Describes me fairly well}\\
\indent \textit{5: Describes me extremely well}\\
\indent \textit{Statement:} \{Survey item\}\\

\indent \textbf{WVS}\\
System prompt:\\
\indent \textit{This is a global study of what people value in life.}\\
\indent \{Persona manipulation\}\\
\indent \textit{Respond only with a number from 1 to 10.}\\
User message:\\
\indent \textit{On a scale of 1 (never justifiable) to 10 (always justifiable), do you think the following action is always justifiable, never justifiable, or something in between:} \{Survey item\}\\

MFQ-2 participants are identified as being from China or the USA, so the MFQ-2 persona prompt states “You are from China”, “You are from the USA”, or is left empty. WVS participants are identified as Chinese or American nationals, so the WVS person prompt states “You are a Chinese national”, “You are an American national”, or is left empty. In the Chinese language condition, all aspects of the prompt are in Chinese.

\begin{figure*}
    \centering
    \includegraphics[width=1\linewidth]{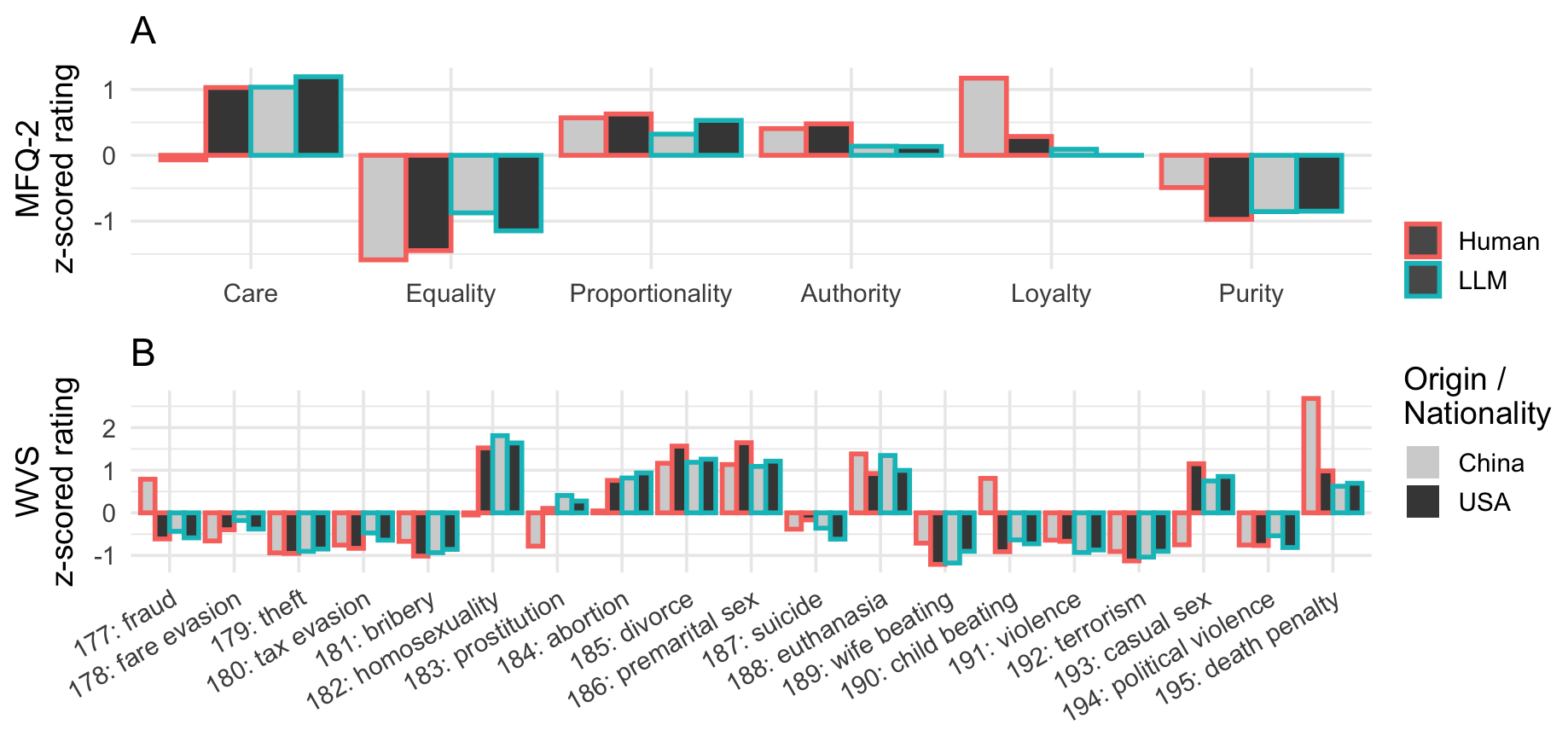}
    \caption{Normalized human and LLM ratings, grouped by dimension for MFQ-2 and question for WVS. Responses are z-scored within each participant and within each LLM. These data exclude the manipulation of persona and language.}
    \label{fig:placeholder4}
\end{figure*}

\subsubsection{RQ1}
To answer RQ1, we fit two linear regression models (separately for the MFQ-2 and WVS), regressing the cosine similarity of human--LLM responses on the interaction of participant nationality (Chinese or American) and LLM origin (China or the USA). We sum coded both factors such that being from China is the positive value (i.e., Chinese = 0.5, American = -0.5). Consequently, a positive main effect of LLM origin would indicate that Chinese LLMs respond more like human participants, and a positive main effect of participant nationality would indicate that all LLMs respond more like Chinese participants. A positive interaction would indicate that the tendency to respond more like Chinese participants is greater in Chinese LLMs than in American LLMs. For the first analysis, we did not manipulate persona, and we paired human responses with LLM responses in the same language. For the WVS, five LLMs (Claude 4, DeepSeek-V2, ERNIE-4.5, GPT-4, and GPT-4o) refused to respond to some items in all 20 iterations, such as whether abortion is justifiable, excluding them from this analysis of similarity on the WVS.

\subsubsection{RQ2}
To answer RQ2, we coded the congruence of participant nationality with LLM country of origin, persona, and language. For example, American participants are congruent with Llama 4 (and the nine other LLMs made in the USA), with the American persona, and with the English surveys. We regressed cosine similarity on the interaction of those three factors, sum coded such that congruence with participant nationality is the positive value (i.e., congruent = 0.5, incongruent = -0.5). For example, a positive main effect of nationality--language congruence would indicate that LLMs respond more like Chinese participants when given a survey in Chinese and more like American participants when given it in English. Positive interactions would indicate that combinations of congruent factors increase similarity more than factors in isolation. When assigned a Chinese or American persona, more LLMs are willing to respond to the WVS (e.g., they state whether abortion is justifiable), so only DeepSeek-V2, GPT-4, and Yi-34B are excluded from this analysis.

\begin{figure*}
    \centering
    \includegraphics[width=1\linewidth]{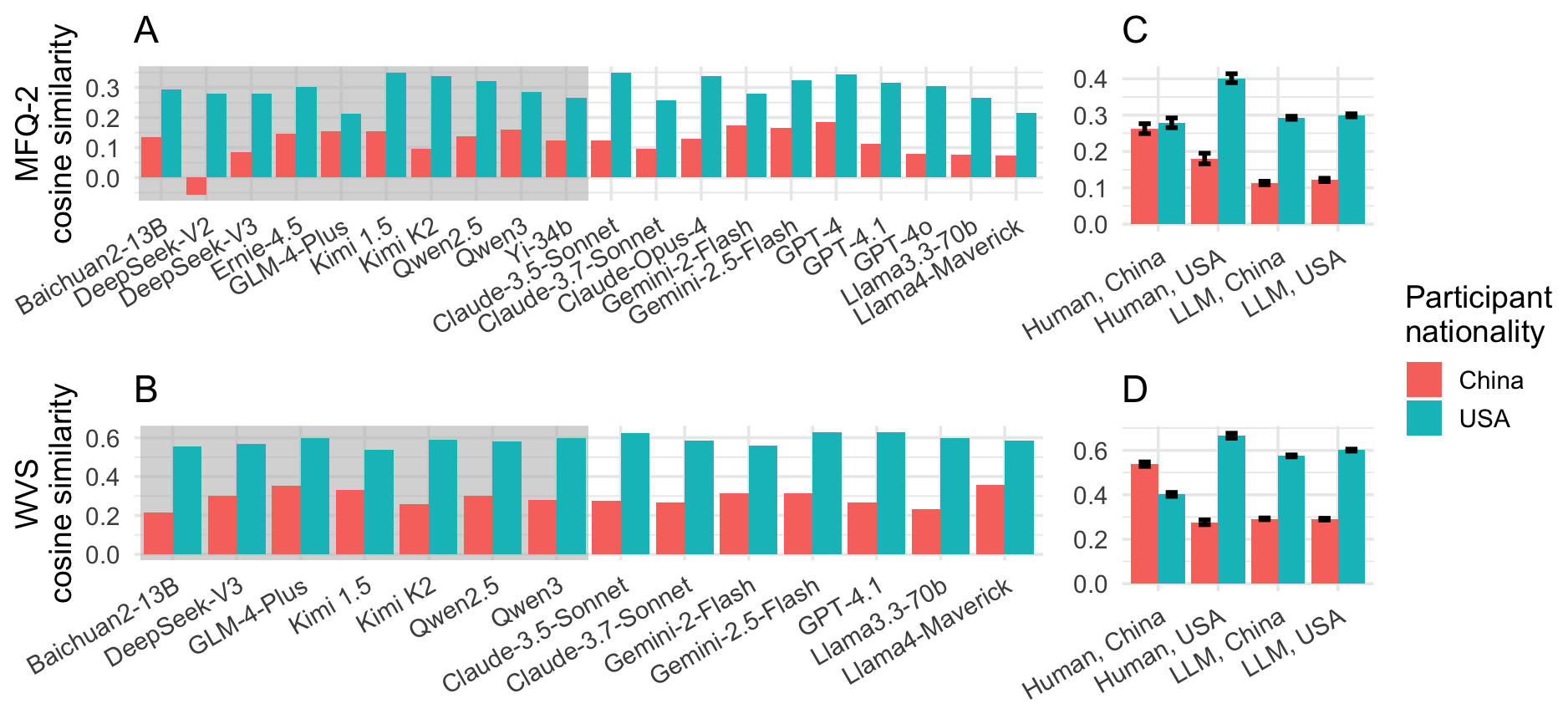}
    \caption{Cosine similarity of individual humans’ ratings to each LLM’s ratings (A and B), and cosine similarity of individual humans’ ratings to mean ratings for each item from LLMs and humans, grouped by country of origin / nationality (C and D). In A and B, the bars with grey backgrounds are LLMs made in China. In C and D, “Human, China” and “Human, USA” refer to the mean ratings to each item from that participant group. “LLM, China” and “LLM, USA” refer to the mean ratings to each item from LLMs made in China and the USA.}
    \label{fig:placeholder0}
\end{figure*}

\subsubsection{RQ3}
 To answer RQ3, we recomputed the cosine similarity of LLM and human response vectors when removing each dimension (for MFQ-2) or each question (for WVS). As with RQ1, we did not manipulate persona, and we paired human responses with LLM responses in the same language. (As we note in Section 4.3, including the persona and language manipulations does not change the direction or significance of these effects.) We regressed the cosine similarity of human and LLM ratings on the interaction of participant nationality, LLM country of origin, and dimension or question removed. We again sum coded nationality and origin such that being from China is the positive value (i.e., American = -0.5; Chinese = 0.5), and crucially, we treatment coded dimension / question such that none removed is the reference level (i.e., none removed = 0; dimension / question removed = 1). The main effects of nationality and origin therefore do not involve removing any dimensions / questions, and the interactions with a given dimension / question indicate how those main effects change when removing that dimension / question. For example, if LLMs tend to be more similar to American participants, nationality will have a negative main effect, but if the Purity dimension were the sole cause of this greater similarity to Americans, then removing Purity would have a positive interaction as large as the negative main effect, canceling it out.\\
 

\section{Results}

Figure 2 illustrates mean human and LLM ratings to the six MFQ-2 moral dimensions and the 19 WVS ethical questions, grouped by nationality for human participants and grouped by country of origin for LLMs. For example, on the MFQ-2, Chinese participants (the light grey bars with red outlines) gave high ratings to questions probing the Loyalty dimension, and on the WVS, they rated the death penalty (Question 195) as very justifiable.

\begin{figure*}
    \centering
    \includegraphics[width=1\linewidth]{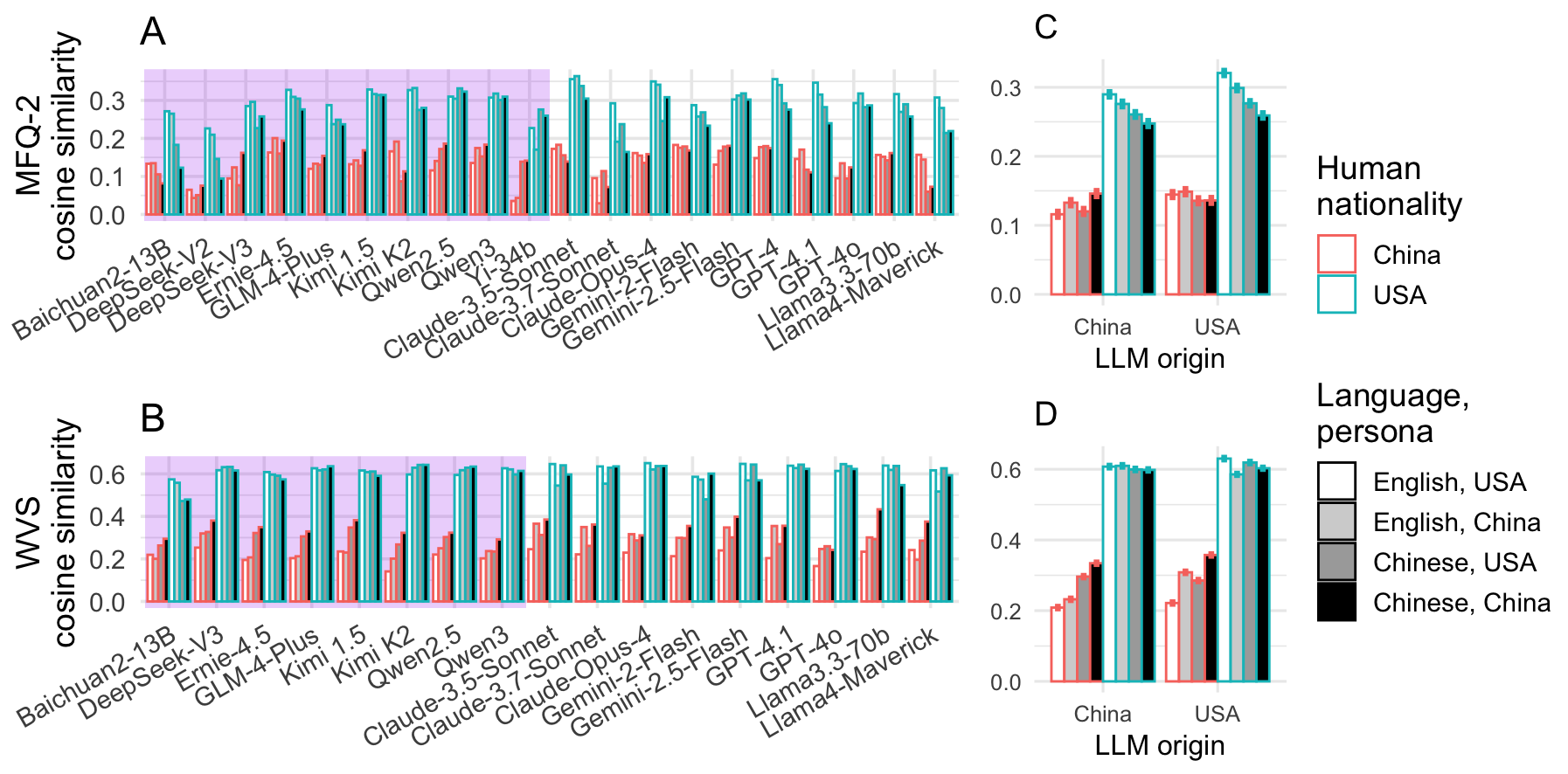}
    \caption{Cosine similarity of LLMs to humans, grouped by LLM country of origin, prompt language, and prompt persona. The bars indicate means, and error bars indicate 95\% confidence intervals. In A and B, the bars with purple backgrounds are LLMs made in China.}
    \label{fig:placeholder2}
\end{figure*}

\subsection{RQ1: Do people and LLMs from the same country share values?}

To investigate whether LLMs from China versus the USA respond more like participants from China versus the USA, we regressed the cosine similarity of each participant’s ratings to each LLM’s rating on the interaction of participant nationality with LLM country of origin. As reported in Table 2, there is a significant negative main effect of participant nationality for both the MFQ-2 and WVS, indicating greater similarity of LLMs to American participants than to Chinese participants (since Chinese participants are coded with the positive value). For the MFQ-2, the interaction of participant nationality with LLM country of origin is far from significant (\textit{p} > .7), therefore providing no evidence that Chinese LLMs tend to respond more like Chinese participants or that American LLMs respond more like American participants. For the WVS, that interaction is significantly positive, indicating that the tendency to respond more like American participants is reduced in LLMs made in China. However, the interaction ($\beta$ = .03) is an order of magnitude smaller than the main effect of participant nationality ($\beta$ = -.30). As is clear in Figure 3, country of origin does little to moderate the greater similarity of LLMs to American participants.\\
\begin{table}[!b]
\centering
\begin{tabular}{lrrr}
\toprule
Predictor & $\beta$ & \textit{SE} & \textit{p} \\
\midrule
\multicolumn{4}{l}{\textbf{MFQ-2}} \\
Intercept & .207 & .001 & $< 2e{-}16$ \\
Language & .017 & .001 & $< 2e{-}16$ \\
Persona & .014 & .001 & $< 2e{-}16$ \\
LLM origin & .004 & .001 & .0012 \\
Language : Persona & .003 & .002 & .2977 \\
Language : Origin & .016 & .002 & $3e{-}11$ \\
Persona : Origin & .013 & .002 & $3e{-}7$ \\
Lang.\ : Pers.\ : Origin & .008 & .005 & .1003 \\
\midrule
\multicolumn{4}{l}{\textbf{WVS}} \\
Intercept & .430 & .001 & $< 2e{-}16$ \\
Language & .042 & .001 & $< 2e{-}16$ \\
Persona & .038 & .001 & $< 2e{-}16$ \\
LLM origin & .008 & .001 & $5e{-}13$ \\
Language : Persona & .006 & .002 & .0104 \\
Language : Origin & .011 & .002 & $6e{-}7$ \\
Persona : Origin & -.014 & .002 & $9e{-}12$ \\
Lang.\ : Pers.\ : Origin & .031 & .004 & $7e{-}13$ \\
\bottomrule
\end{tabular}
\caption{A linear regression model regressing the cosine similarity of LLM--human responses on the congruence of participant nationality with LLM country of origin, survey language, system prompt persona, and their interactions. All factors are sum coded and mean centered with congruent as the positive value.}
\end{table}
\indent As a reference point, Figures 3C and 3D contrast the similarity of humans to LLMs with the similarity of humans to humans. We calculated the mean rating to each item from Chinese LLMs and American LLMs and the mean rating to each item from Chinese participants and American participants. Then, we computed the cosine similarity of those mean ratings to individual participants’ ratings. On the WVS (Figure 3D), mean ratings from Chinese participants are more similar to individual Chinese participants’ ratings than to individual American participants’ ratings, as expected. However, on the MFQ-2 (Figure 3C), mean Chinese ratings are instead slightly more similar to individual American participants’ ratings. To confirm that LLMs respond significantly more like American participants than Chinese participants do, we conducted a follow-up analysis. As shown in the online supplement, we compared the similarity of individual participants’ ratings to mean Chinese ratings versus the similarity of individual participants’ ratings to each LLM’s ratings. The effect of participant nationality is substantially greater for each of the 20 LLMs than for mean Chinese ratings. So, on the MFQ-2, Chinese participants tend to respond slightly more like American participants than like other Chinese participants, on average, which underscores how values vary within a society, but the greater similarity to American participants is far more pronounced in LLMs than in Chinese participants.

\begin{figure*}
    \centering
    \includegraphics[width=1\linewidth]{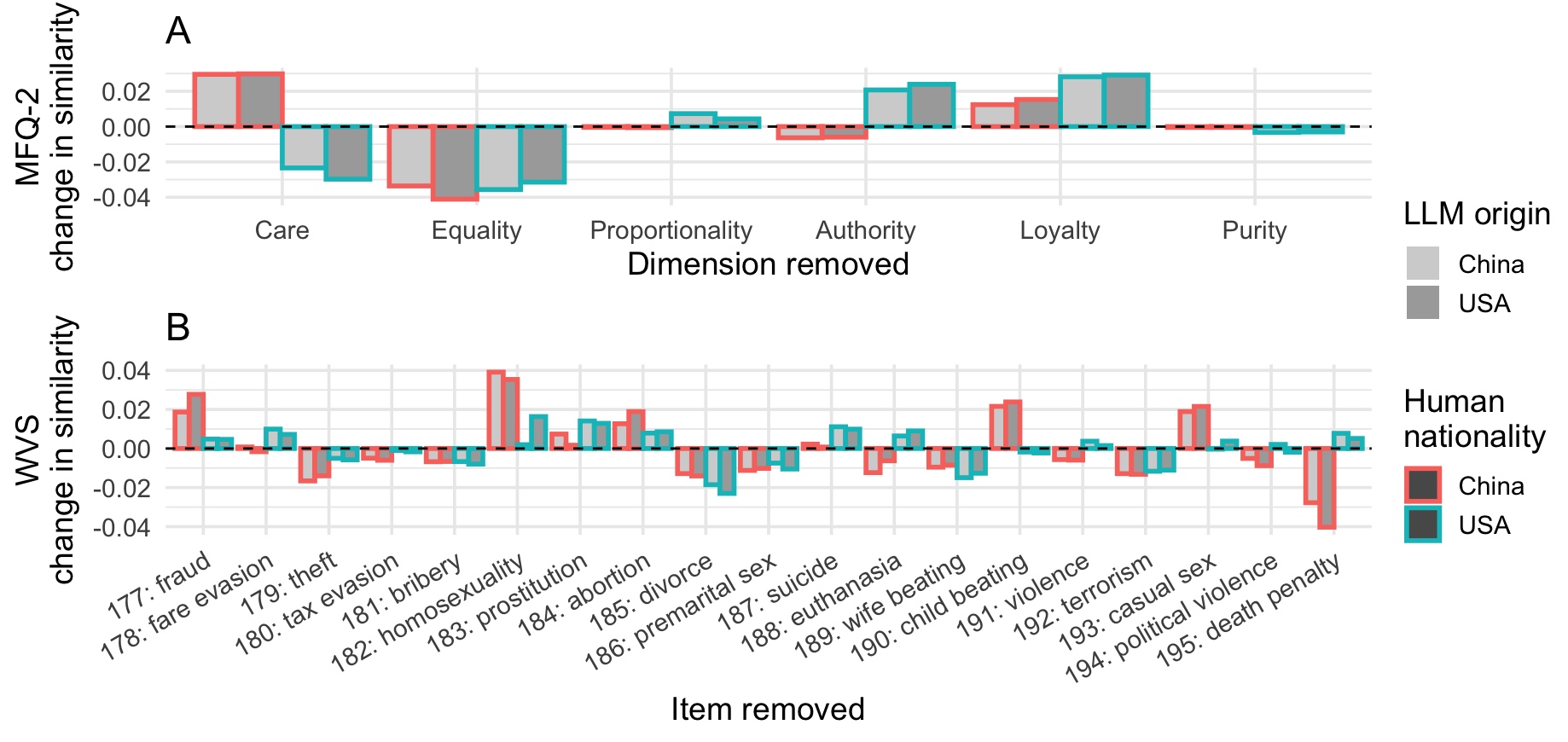}
    \caption{Change in cosine similarity of LLMs to humans when removing a dimension (MFQ-2) or item (WVS).}
    \label{fig:placeholder3}   
\end{figure*}

\subsection{RQ2: Does language or persona steer LLMs toward human populations?}

Next, we investigated whether manipulating prompt language and/or imposing personas on LLMs made their responses more like those of the corresponding participant nationality. We coded the congruence of participant nationality with prompt language, imposed persona, LLM country of origin, and their interactions. For example, Chinese participants are congruent with Chinese prompts, Chinese personas, and LLMs made in China. Then, we regressed the cosine similarity of individual participant ratings to LLM ratings on the interaction of those congruence factors. As reported in Table 3, for both the MFQ-2 and WVS, the main effects of nationality–language congruence, nationality–persona congruence, and nationality–origin congruence are all significantly positive: The greater similarity of LLMs to American participants is mitigated when LLMs respond to surveys in Chinese, are told they are from China, and are made by a company from China. For the WVS, the positive main effect of nationality–origin congruence is consistent with the significant positive interaction of participant nationality with LLM origin reported in Section 4.1. However, for the MFQ-2, the positive main effect of nationality–origin congruence differs from the null interaction of participant nationality with LLM origin reported in Section 4.1: In the current analysis, the skew toward American ratings is reduced in Chinese LLMs, whereas in Section 4.1, we found no evidence that being made in China reduces that skew. Recall that Section 4.1 did not impose personas on LLMs, so this discrepancy demonstrates how adding persona prompts changes LLM behaviour.\\
\indent The main effects of congruence are small but consistent, whereas combinations of congruent factors have mixed effects. For the MFQ-2, the interactions of nationality–origin congruence with nationality–persona congruence and with nationality–language congruence are both significantly positive, indicating that the congruent persona and the congruent language more readily steer LLMs toward responses of participants from the same country as the LLM. In Figure 4C, this can be observed in the boost for American nationality + USA persona bars for American LLMs and in the Chinese nationality + China persona bars for Chinese LLMs. For the WVS, the interaction of origin with persona is instead significantly negative, indicating that, on average, personas are better at steering LLMs toward participants from the opposite country. In Figure 4D, this can be observed in the Chinese nationality + China persona bars, which exhibit a slightly greater boost in American LLMs than in Chinese LLMs. However, further complicating matters, the three-way interaction is significantly positive for the WVS, suggesting that the congruent survey language helps LLMs adopt the congruent persona. The key takeaway, evident in Figure 4, is that although these factors steer LLMs toward the expected human populations, they do little to mitigate the greater similarity of LLMs to American participants.

\subsection{RQ3: Which items align LLMs with American participants?}
 
Finally, we investigated which MFQ-2 dimensions and WVS questions cause the overall greater similarity of LLMs to American participants. We recomputed the cosine similarity of human ratings to LLM ratings when removing the six items that correspond to each MFQ-2 dimension or when removing each WVS question. In two linear regression models, reported in full in the online supplement, we regressed the cosine similarity on the interaction of LLM country of origin, participant nationality, and the removed dimension / question. For the MFQ-2, Authority and Loyalty have significantly positive main effects ($\beta$s = .008 and .021, \textit{p}s = 1e-6 and < 2e-16, respectively), indicating that removing those dimensions increases the similarity of LLMs to humans, regardless of nationality. Equality has a significantly negative main effect ($\beta$ = .035, \textit{p} < 2e-16), so removing it decreases similarity. 
More importantly, Authority and Loyalty have negative interactions with participant nationality ($\beta$s = -.028 and -.015, \textit{p}s < 2e-16 and = 1e-5, respectively), indicating that removing those dimensions decreases similarity to Chinese participants relative to American participants (because Chinese nationality is sum coded as the positive value). Explaining the greater similarity to American participants instead requires a significant positive interaction (i.e., removing that dimension increases similarity to Chinese participants), and only the Care dimension meets this requirement. The interaction of Care with nationality ($\beta$ = .056, \textit{p} < 2e-16) is by far the largest of any effect other than the main effect of participant nationality ($\beta$ = .175). This is apparent in Figure 5 and, above, in Figure 2, where Chinese participants are conspicuously dissimilar to LLMs and American participants on the Care dimension. As reported in full in the online supplement, the interaction of Care with LLM country of origin is far from significant (\textit{p} = .45), indicating that the inclination toward high Care ratings does not differ in LLMs from China versus the USA.\\
\indent To explain the greater similarity of LLMs to American participants on the WVS, we again looked for large positive interactions of question with participant nationality (relative to the main effect of participant nationality: $\beta$ = .278). Four questions fit the bill: Q177 (taking illicit government benefits; $\beta$ = .018, \textit{p} < 2e-16), Q182 (homosexuality; $\beta$ = .028, \textit{p} < 2e-16), Q190 (corporal punishment; $\beta$ = .025, \textit{p} < 2e-16), and Q193 (casual sex; $\beta$ = .019, \textit{p} = 8e-11). As illustrated in Figure 2, these questions differentiate Chinese participants from American participants, such that Chinese participants find homosexuality and casual sex less justifiable and find taking benefits and corporal punishment more justifiable. We also found smaller but significant positive interactions with participant nationality for Q184 and Q185 (abortion and divorce, respectively; $\beta$s = .008 and .007, \textit{p}s =.008 and .010, respectively), which Chinese participants rate as less justifiable than American participants. Only Q182 (regarding homosexuality) has a significant three-way interaction with participant nationality and LLM country of origin ($\beta$ = .018, \textit{p} = .001). The positive effect indicates that removing Q182 increases similarity to Chinese participants more for Chinese LLMs than American LLMs, the implication being that Chinese LLMs deem homosexuality slightly more justifiable, on average, consistent with Figure 2. Because these null persona data exclude Claude 4, DeepSeek-V2, ERNIE-4.5, GPT-4, and GPT-4o, we conducted a follow-up analysis using the LLM persona manipulation (which excludes only DeepSeek-V2, GPT-4, and Yi-34B, as in Section 4.2), and we found the same pattern and significance of effects.

\section{Discussion}

We measured the similarity of LLM responses to human responses on the MFQ-2 and the seventh wave of the WVS. We found that all 20 LLMs responded more like American participants than like Chinese participants. This skew toward American values was only slightly mitigated when LLMs were made in China, when imposing a Chinese identity on the LLMs, and when presenting LLMs with surveys and prompts in Chinese. On the MFQ-2, the greater similarity to Americans can be traced back to the Care dimension, where American participants and LLMs gave substantially higher ratings than Chinese participants did. On the WVS, the greater similarity to Americans can be traced back to a few questions: LLMs and Americans rate homosexuality and casual sex (and to a lesser extent, abortion and divorce) as more justifiable than Chinese participants do, and they rate corporal punishment and taking illicit government benefits as less justifiable than Chinese participants do. Bear in mind that these differences emerge only in the aggregate and do not reliably predict individuals’ values.\\
\indent Why do LLMs tend to respond more like American participants than like Chinese participants? Secrecy surrounds state-of-the-art LLMs---companies that tout themselves as open source, such as Meta and DeepSeek, do not reveal their training data, and OpenAI has not even released the embedding sizes or number of layers in their LLMs---so we can only speculate as to the causes of LLM behaviour. Training data is probably one important factor. OpenAI has alleged that DeepSeek relied on training data distilled from GPT models (\citealp{sweney2025}), consistent with anecdotal evidence that DeepSeek identifies itself as GPT-4 (e.g., \citealp{wiggers2025}). Training on the output of big, slow, and expensive LLMs can allow companies to train smaller, faster, and cheaper LLMs with competitive performance \cite{hsieh2023distilling}, and one consequence of this distillation might be that smaller “student” LLMs absorb the values of larger “teacher” LLMs. However, this begs the question as to why the teacher LLMs skew toward American values in the first place. Again, training data is probably a factor. According to estimates from the Common Crawl, about half of the text on the internet is written in English \cite{commoncrawl}, and given the need to train LLMs on as much text as possible, this availability bias may impose English speakers’ values on LLMs.\\
\indent Fine-tuning probably plays an important role, too. Before pretrained models are released to the public, they undergo reinforcement learning to align them with human preferences (e.g., \citealp{bai2022training}). A standard principle in reinforcement learning is that LLMs should be helpful, honest, and harmless \cite{askell2021general}, and on the MFQ-2, Harm is the flip side of Care \cite{haidt2007morality}. Recall that American participants and LLMs gave higher ratings to items from the Care dimension than to any other dimension (i.e., they reported that pro-Care statements described them best), whereas the Chinese participants gave only average ratings to those items. There is, then, a straightforward argument that stressing harmlessness during fine-tuning aligns LLMs more with American people’s responses to the MFQ-2 than with Chinese people’s responses.\\ 
\indent However, harmlessness may not account for the alignment of LLMs with American participants on the WVS, and insofar as it does account for that alignment, it reveals normative assumptions about what constitutes harm. LLMs have been shown to align with liberal values (e.g., \citealp{hartmann2023political, motoki2024more, rozado2024political}), and their responses regarding homosexuality and corporal punishment are consistent with a liberal perspective (e.g., \citealp{graham2009liberals}). In this way, the seemingly uncontroversial principle of harmlessness delays a confrontation with the paradox of tolerance \cite{popper1945}: When LLMs rate corporal punishment as unjustifiable, for instance, they demonstrate how liberal ideals prohibit behaviours that some people believe to be valuable. The relevant point for the purpose of our study is that this liberal, American perspective persists in LLMs made in China. That could be a conscious decision made by Chinese AI companies, but more likely, it reveals how standard training methods and available training data perpetuate the values held by a narrow slice of humanity. As LLMs become pervasive, it will be increasingly important that users understand what values LLMs tolerate and propagate.

\section{Conclusion}

Our findings show that, despite Beijing’s directive to align AI systems with “socialist core values,” \cite{govcn2023}, LLMs made in China tend to reinforce American values. If Chinese LLMs inherit those values from predominantly English corpora and from training methods designed in the USA, China may struggle to project its own norms through AI. This suggests that the AI may become a new channel for exporting American liberal norms.

\section{Limitations}

This study operationalizes moral values as how people and LLMs respond to two widely used surveys, the MFQ-2 and WVS. Future work should investigate whether different surveys and different approaches to eliciting moral values corroborate our findings. This study selected 10 LLMs from each of the USA and China, based on performance and availability. This is a fast-moving field, so future work should investigate whether our findings hold across other LLMs.

\bibliography{bibtex_DAVE_EDIT}
\bibliographystyle{acl_natbib}
\end{document}